# P$^3$FA: Unified Unicast/Multicast Forwarding with Low Egress Diversities

Zhu Jin and Wen-Kang Jia, *Senior Member, IEEE*

*Abstract*—Multicast is an efficient way to realize one-to-many group communications in large-scale networks such as the Internet. However, the deployment of IP multicast services over the Internet has not been as rapid as expected and needed. Excepting the fatal defects in designing IPv4 address structure. Another main reason that contributes to this slow deployment is the lack of carrier-grade multicast-enabled switches and routers that can be as to scale as their unicast counterparts. Implementing a high-performance switch/router relies on a polynomial-time group membership query algorithm within the Packet Forwarding Engines (PFEs) to determine whether or not a packet is forwarded through an egress. Among these, Bloom filter (BF)-based and Residue Number System (RNS)-based are being considered as two representations of the membership query algorithms. However, both approaches suffer from some fatal weaknesses such as space and time inefficiencies, especially for a carrier-grade PFE with high port-density features. According to similar properties of the prime theorem, we propose a simplified forwarding scheme in this paper, named Per-Port Prime Filter Array (P$^3$FA). The simulation results indicate that the P$^3$FA can significantly improve space efficiencies under specific lower egress-diversities conditions. Under the same space constraints, compared with the SVRF, the multicast time efficiencies, the unicast time efficiency of the P$^3$FA are respectively increased by 12x-17234x and 19x-2038x at a range of port-densities 16-1024, but at the expense of hardware cost, which increased by $\rho$/2x. A PFE designer that attempts to adopt P3FA should trade-off between required performance and cost.

## I. INTRODUCTION AND PROBLEM STATEMENTS

The multicast protocol is developed to deliver real-time data to multiple receivers efficiently over large-scale networks, and it becomes increasingly important for various applications such as video-conferencing, video on demand, and live streaming, etc. nowadays [1]. Traditional IP multicast schemes are based on the efficient management of one-to-many network resources, which can deal with very large multicast groups, but they may encounter scalability problems with a very large number of distinct multicast groups. Modern Internet applications frequently exhibit one-to-many communication patterns and, at the same time, require sub-millisecond latencies and high throughput. IP multicast can achieve these requirements but has control- and data-plane scalability limitations that make it challenging to offer it as a service for hundreds of thousands of tenants, typical of internet environments [2].

As a core component of carrier-grade (large-scale) packet routing/switching equipment, multicast-enabled Packet Forwarding Engine (PFE) has been widely used in some fields and is still under development. Inside a PFE, similar to unicasting, the multicast routing table (MRT) is usually converted into a multicast forwarding table (MFT) in most cases, which contains all multicast forwarding entries (MFEs), and the MRT is generally much smaller than the MRT [3]. In addition to a unique flow identifier, e.g., multicast address, an MFE also includes the multiple output port identifiers and the next-hop node identifiers. Under normal circumstances, a relatively higher efficient multicast forwarding algorithm will directly control the data-plane (switch fabric) to forward the multicast packets/flows alone the on-tree switches/routers according to the MFT [3].

Since an MFE must represent more complicated behaviors, which requires huge amounts of memory resources to store the huge amount of extra states includes its own specific branch pattern (i.e., indicating that it has plural egresses), which cannot be simply aggregated by prefix as in the unicasting [4][5]. The available member space of multicast routing table (MRT)/ multicast forwarding table (MFT) may fill up quickly when a certain number of multicast groups that launched on the network. As is well-known that the addressing space of IPv4 is equipped with $2^{28}$ multicast addresses, and the IPv6 is equipped with up to $2^{120}$ multicast addresses. It is expected that a serious unscalable issue will occur at each multicast-enabled switch/router on the Internet and often attributed as a structural bottleneck in the large-scale networks that involve multicasting [6].

Therefore, to meet the increasing demand for high-performance computing and connect thousands of upper-layer network devices and end-hosts, scalability and efficiency have become important performance indicators for today's core IP routers and high-performance switches. A PFE's forwarding performance is mainly depending on its multiset membership query algorithm, which is a variation of membership query algorithm, it can provide a consistent view of the multiset group membership query process, and offer ultra-efficient forwarding performance of PFE.

Besides the well-known Bloom filter (BF) [7][8] was widely adopted as a group membership query in modern carrier-grade PFE. Another approach—SVRF [9] [10] mainly benefits from the Residual Number System (RNS) [11], which can obtain egress information from a large public scalar pair in pseudo-polynomial time through a unique key corresponding to the data flow, which has achieved good results. The SVRF scheme is consisting of 1) a pair of scalar $M_{CP}$ and $M_{CRT}$, which stored in two independent memory blocks; 2) two unsigned integer dividers; and 3) a prime generator, which maps each packet identifier toward a unique member-specific key. When an incoming packet arrives at the PFE, the element identifier (e.g., destination address) is extracted, then mapped (or hashed) to a unique member-specific key. In mathematics, the 1$^{st}$ scalar $M_{CP}$ is the continuous product (CP) of all key∈G that cannot be divided exactly by any keys∉G, thus any keys are the result of either hit (zero) or miss (non-zero) values. Once hit, based on the properties of the Chinese remainder theorem (CRT) [12], the 2$^{nd}$ scalar $M_{CRT}$ that divided by different keys∈G will remain as different remainders as vectors. We use these vectors to represent the exact output port index (OPI) and output port bitmap (OPB) in unicast and multicast respectively. Therefore, the SVRF can simply obtain the corresponding OPI/OPB of each multicast group by dividing a common scalar-pair (stored in memory) with a group-specific key (mapped from packet identifier) within pseudo-polynomial time. As an error-free group membership query algorithm that allows insertion and deletion operations easily, SVRF would provide a solution that met the main requirements of the multicast forwarding algorithm. However, when the port-density of the PFE is high, a large number of prime numbers multiply will produce larger scalar-pairs, which will cause the memory space occupied by SVRF to increase exponentially. Consequently, SVRF faces the problem of scalability under finite memory constraints. In addition, the division of very large integers will consume a lot of computation time, resulting in low time efficiency, even if used hardware acceleration is also difficult to improve significantly. It can be said that the SVRF sacrifices time efficiency for space efficiency, and its query performance is basically poor than in BF approaches.

Based on the identical concept of SVRF, the Fractional-$N$ SVRF [10] scheme was proposed as an improved alternative solution. In order to reduce the memory space of SVRF and optimize time efficiency, Fractional-$N$ SVRF randomly divides $n$-element forwarding entries into $N$ groups to generate $N$ relatively small scalar pairs to replace the large scalar pairs generated by SVRF. Each group is equivalent to an independent SVRF sub-block and performs query operations independently, hence when element's keys belonging to distinct sub-blocks are allowed to reuse relatively smaller identical prime keys. Thus, it is expected that $N$ smaller scalar-pairs can execute query tasks in parallel, which can effectively reduce both memory consumption and computational complexity to obtain more high forwarding performance.

Considering that most multicast packet flows belonging to certain multicast groups will not be forwarded toward a relatively high number of distinct egress in a PFE. On the contrary, with specific topological models, most multicast group users are spread over a specific and small number of domains. As a result, the desired egress in a given multicast forwarding entry should be sparsely represented in the OPBs in a PFE. For example, a multicast packet might be forwarded to a single egress or relatively few egresses instead of a relative majority of egresses. In other words, this multicast delivery tree is not branched over large amounts of egresses in this switch/router node. We called this situation the low egress-diversity of the forwarding node. The egress-diversity $\varphi$ refers to the average number of concurrent egresses of all MFEs in MFT during a particular time interval. We can periodically count the average number of concurrent egresses of all MFEs by

$$\varphi = \sum_{i=1}^{n} \langle OPB_i \rangle \Big/ n, \quad (1)$$

where $n$ denotes the number of MFEs, and bit counter function $\langle OPB_i \rangle$ represents the number of bit-1 of the OPB of the $i$-th MFE in MFT. For example, the unicasting is with the $\varphi=1$ and the multicasting is with the $\varphi=1\sim\rho$. In such a low egress-diversity situation, the SVRF utilizes scalar-pair ($M_{CP}$, $M_{CRT}$) to deal with a few bit-1 even single bit-1 OPBs, which seems to be lacking efficiency on memory space. In this paper, we discuss a new approach, called Per-Port Prime Filter Array (P³FA), in which the main idea is to further reduce the total memory usages and forwarding lookup latencies of the SVRF under specific conditions such as lower egress-diversities.

## II. PER-PORT PRIME FILTER ARRAY (P³FA)

Due to the aforementioned reasons, we propose a simplified P³FA scheme to further reduce both the memory consumptions (space) and lookup latencies (time) in this section.

### A. The P³FA Construction

The function block diagram with major components of a P³FA is shown in Fig. 1, The major components in a P³FA include the 1) input and output units, the 2) P³FA processing unit, and the 3) memory unit, the features of which are described as follows. The control-plane of PFE carries out the instructions of forwarding direction (egress(es)), which are given by the upper-layer component such as routing engine (RE) implementing by dynamic routing protocols and Routing Information Base (RIB). The Parser is to extract the incoming data flow to obtain the flow identifier. The function of the Prime Hash is to assign a unique $\rho$-bit prime key to the data flow by identifying the flow identifier. Each port is corresponding to a very-long integer 'Divider', it is mainly used to calculate the modulus value of $M_{CP}$ and key, where $M_{CP}$ stores at the memory unit, as the dividend; the key is generated from 'Prime Hash' and as the divisor. The result of the calculation is the $\rho$-remainders, which is either zero or non-zero, mainly used to determine whether the packet is forwarded to this port. With a combination of a '$\rho$-Inverter Array' and a '$\rho$-DEMUX (demultiplexer)', which combinational logic circuit is mainly responsible for enabling all dividers except the $x$-th Divider$_x$ (Disabling Divider$_x$), where $x$ denotes the ingress ID.

Within the proposed P³FA, we simply separate the one-big group membership into $\rho$ partitions, which $\rho$ denotes the port-density (number of the logical interfaces) of the PFE. Hence we have several relatively lesser sub-scalar arrays $\{M_{CP1}, M_{CP2}, …, M_{CP\rho},\}$ instead of one-big scalar-pair $M_{CP}$ in the memory unit. The processing unit of the SVRF is also separated into $\rho$ sub-blocks in P³FA, and each sub-block is associated with each interface (potential egress). Based on the properties of the prime theorem, an $M_{CP}$ divided by different keys will remain as zero or nonzero results, which can be used to answer inquiries regarding whether one element $k_i$ (associated to packet/flow $i$) belongs to specific egress(es).

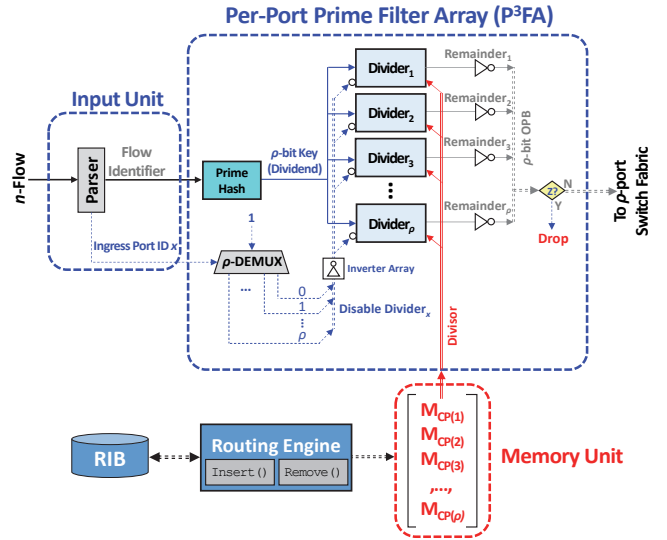

Fig. 1. The block diagram of the P³FA functionality within the $\rho$-port PFE.

The egress-diversity $\varphi$ is an index value indicating the distinct egresses of total egresses that a given MFT, which forwards multicast packets with a distribution pattern per unit time. For example, a typical multicast packet in an intermediate switch/router node forwards to a relatively restricted number of egress, whereas a broadcast frame may utilize the total egresses. Therefore, when the egress-diversity $\varphi$ is low enough, it might satisfy

$$\left( m_{p3fa} = \sum_{s=1}^{\rho} \|M_{CP(s)}\| \right) < \left( m_{svrf} = \|M_{CP}\| + \|M_{CRT}\| \right), \quad (2)$$

where $\|M_{CP(s)}\|$ denotes the bit-length of the sub-scalar $M_{CP(s)}$ corresponding to each egress $s$, and $m_{svrf}$ denotes the bit-length sum of the sub-scalar $M_{CP}$ and $M_{CRT}$ of SVRF.

### B. Constructing the P³FA Scalar-matrix

Assuming that we have a $\rho$-port PFE with supporting maximum membership capacity $n$. Traversing all the entries from MFT, we obtain the corresponding OPB array $B=\{b_1, b_2, ..., b_n\}$ of each membership. Each OPB $b_x$ contains $\rho$ bits, that is, $b_x=\{b_{x(1)}, b_{x(2)}, …, b_{x(\rho)}\}$. The value of $b_{x(s)}$-bit indicates whether the data flow is forwarded through the port $s$: a bit-0 value indicates that the flow should not be forwarded along with the port, while a bit-1 value indicates that the flow should be forwarded along with the port. Then each data flow will be randomly assigned a key, we have $K=\{k_1, k_2, ..., k_n\}$, each $k_x$ (an integer) is a unique prime number with a one-to-one correspondence with $b_x$. The sub-scalar array $\{M_{CP(1)}, M_{CP(2)}, …, M_{CP(\rho)}\}$ then constructed from

$$\left[ M_{CP(1)}, M_{CP(2)}, ..., M_{CP(\rho)} \right] = \prod_{x=1}^{n} \left[ k_x^{b_{x(1)}}, k_x^{b_{x(2)}}, ..., k_x^{b_{x(\rho)}} \right]. \quad (3)$$

In the prime number theorem, a sub-scalar $M_{CP(s)}$ is the continuous product (CP) (an integer) of all key $k_x \in K$, and that cannot be divided exactly by any key $k_y \notin K$, thus any keys are the result of either hit (zero) or miss (non-zero) values. Dissimilar to the SVRF, the $s$-th sub-scalar $M_{CP(s)}$ only multiplied by the key $k_x$ of the data flow forwarded to the desired $s$-th port, and the values of $k_x$ are no longer need to greater than values of $b_x$. Under the premise of not repeating each other's, the key can be an arbitrary bit prime number, and the relatively smaller the better, such as 2 (the first prime number).

For example, for a 4-port PFE, suppose that we have a current scalar array $M_{CP(1)}=213$, $M_{CP(2)}=3003$, $M_{CP(3)}=1309$, and $M_{CP(4)}=2431$. When a new member (flow) $x$ arrives from port 1 is added, it is assigned a new key $k_x=23$ with OPB $b_x=\{1,1,0,0\}$. Then the scalars will be updated to $M_{CP(1)}=213$, $M_{CP(2)}=3003$, $M_{CP(3)}=23\times1309=30107$, and $M_{CP(4)}=23\times2431=55913$ in the memory unit.

However, when the egress-diversity is high, it means that the greater number of prime keys is generated the greater values of each sub-scalar $M_{CP(s)}$, resulting in an exponential increase in the total memory consumptions. We think that we could decrease memory consumption as much as possible by manipulating the distribution of keys. For example, for specific data flows with a high egress-diversity, a smaller prime number key can be assigned preferentially. Although this manner can save some memory, it is undoubtedly more complicated to construct a scalar matrix, so the feasibility needs to be studied later.

### C. Querying the P³FA Scalar-matrix

Once a packet arrives at an input unit of P³FA, a packet-identifier (e.g., destination IP address) will be first fetched by the parser. Then the prime hash will generate a prime key $k_x$ uniquely corresponding to the packet-identifier. Note that an identical packet-identifier would always be mapped to an identical key. Then, the key $k_x$ will be delivered to the dividers 1, 2, ..., $\rho$ corresponding to all ports for modulo operation except the ingress $x$ (assuming that the input port is disallowing for re-forwarding, the purpose is to prevent loops). Finally, the OPB $b_x$ of the flow packet is obtained by merging the calculation results of each divider and instructed to the switch fabric (crossbar) to complete the forwarding task. In the query phase for an element $x$, its OPB $b_x$ will be obtained by

$$b_x = \left[b_{x(1)}, b_{x(2)}, ..., b_{x(\rho)}\right]$$
$$=\sim\left(\left[M_{CP(1)}, M_{CP(2)}, ..., M_{CP(\rho)}\right] \mod k_x\right), \quad (4)$$

where the '~' denotes the inverse operator, it makes the value of 'non-zero' become bit-0 and the value of 'zero' become bit-1. Note that the completed OPB is calculated apart from each sub-scalar of P³FA, and combined at one.

For example, for a previous 4-port PFE, when a packet belongs flow $x$ is arrival from port 1, the prime number hash function assigns the previous key $k_x=23$ by recognizing its packet-identifier. Then $k_x$ will be copied into 3 copies and sent to the dividers corresponding to ports 2, 3 and 4, and then perform modulo operation with scalar $\{M_{CP(2)}, M_{CP(3)}, M_{CP(4)}\}$ respectively. Based on current $M_{CP(2)}=3003$, $M_{CP(3)}=30107$, and $M_{CP(4)}=55913$, we have ($M_{CP(2)} \mod k_x$)=13 (non-zero), ($M_{CP(3)} \mod k_x$)=0, and ($M_{CP(4)} \mod k_x$)=0. This means that packet $x$ will be forwarded to port 3 and port 4 simultaneously.

Unlike the SVRF, the distributed structure makes the value of each sub-scalar no longer suffer from the complicated RNS modulo calculation on very-large operands when querying the packet flows' membership, thus the forwarding latency in P³FA will also be accelerated on paralleled sub-blocks, thus saving a lot of computation resources. On the other hand, multiple sub-blocks interconnect memory units might through multiple data paths, such structure could have guaranteed for achieving high performance on PFEs. In addition, because the bit-length of the prime key is no longer required to greater than the corresponding OPB, even though the identical key cannot be reused in distinct scalar arrays, it still constructs the relatively smaller scalar possibly, thus improve the space efficiency on PFEs.

### III. PERFORMANCE EVALUATION

To demonstrate the performance of the P³FA and competitive schemes under various port-densities of PFEs, we build a representative PFE and associated membership algorithms using the Maple® 12 mathematic simulator. We manipulate different port-density (typically from 16 to 1024-port) PEF scenarios during the whole simulation. Each port-diversity $\varphi$ instantiates a respectively different number of desire forwarding ports (typically from $2^8$ to $2^{20}$ egress) for the reference MFEs. Typically, the performance of a PFE is evaluated in terms of space and time efficiencies.

### A. Space Efficiency

The space efficiency is expressed as the total number of required memory space $m$ versus the membership capacity $n$ of the observed forwarding schemes. We compare P³FA to the SVRF and Fractional-$N$ SVRF with various parameters such as port densities $\rho$, egress-diversities $\varphi$ are taken into account. Unlike P³FA, SVRF [9] and Fractional-$N$ SVRF [10] both contain two memory scalar pairs $M_{CP}$ and $M_{CRT}$. Because Fractional-$N$ SVRF divides scalars into $N$ sub-blocks for parallel processing, the space and time efficiency is superior to SVRF. In addition, it is worth mentioning that a multicast (OPB) use a relatively longer ($\rho$+1)-bit key per multicast forwarding entry, and a unicast (OPI) only use a relatively shorter $\lceil \log_2(\rho+1)\rceil$-bit key per unicast forwarding entry.

We should understand that the memory consumptions of P³FA should increase as the egress-diversity $\varphi$ increases. However, the memory consumptions of the SVRF have nothing to do with $\varphi$, when the values $n$ and $\rho$ are determined, the $m_{svrf}$ is a constant. To quantify low and high egress diversity, we introduce the egress diversity threshold $\Phi$, where the value of $\Phi$ is preset by us. When the egress diversity $\varphi>\Phi$, it is the high egress diversity. On the contrary, when $\varphi<\Phi$, it is the low egress diversity. Therefore, we need to discuss the performance thresholds $\Phi$ when the $m_{p3fa}$ will exceed the $m_{svrf}$ under different port-density $\rho$.

The simulation results are depicted in Fig. 2(a)~(d) for the distinct number of forwarding entries among $2^8$, $2^{12}$, $2^{16}$, and $2^{20}$, respectively. First, with the fixed values $n$, we can observe that as the $\varphi$ increases, the memory consumptions of P³FA will increase logarithmically. In addition, as the port-density $\rho$ increases, the egress-diversity threshold $\Phi$ (when $\varphi>\Phi$, $m_{p3fa}>m_{svrf}$) will increase accordingly. Then, as the $n$ increases, the memory consumptions of the three schemes will increase exponentially. It is worth noting that when the port-density $\rho$ is the same, the egress-diversity threshold $\Phi$ decreases as the number of forwarding entries $n$ increases. The reason is that as the $n$ increases, the memory consumptions of P³FA increases while the memory consumptions of SVRF are almost unchanged.

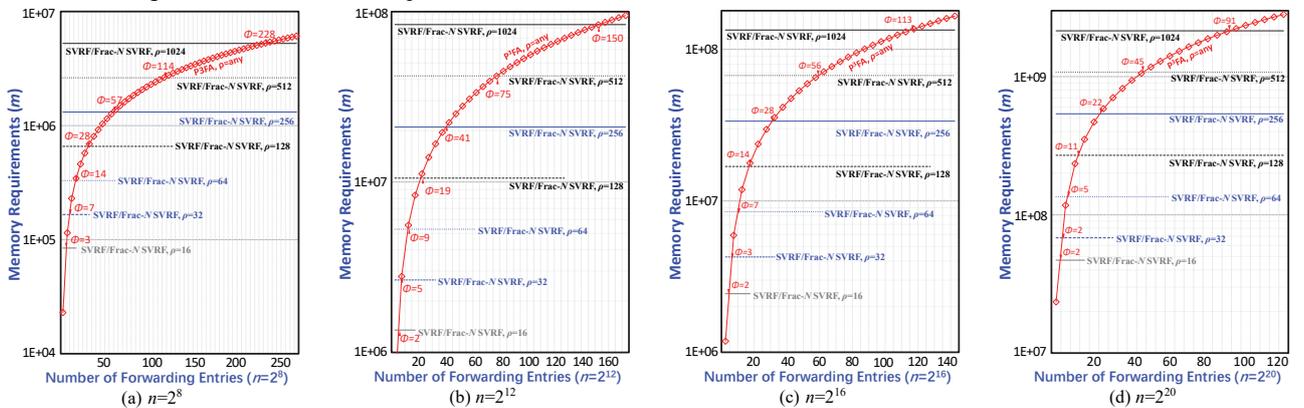

Fig. 2. Required memory space $m$ versus the egress-diversities $\varphi$ in the number of forwarding entries $n$.

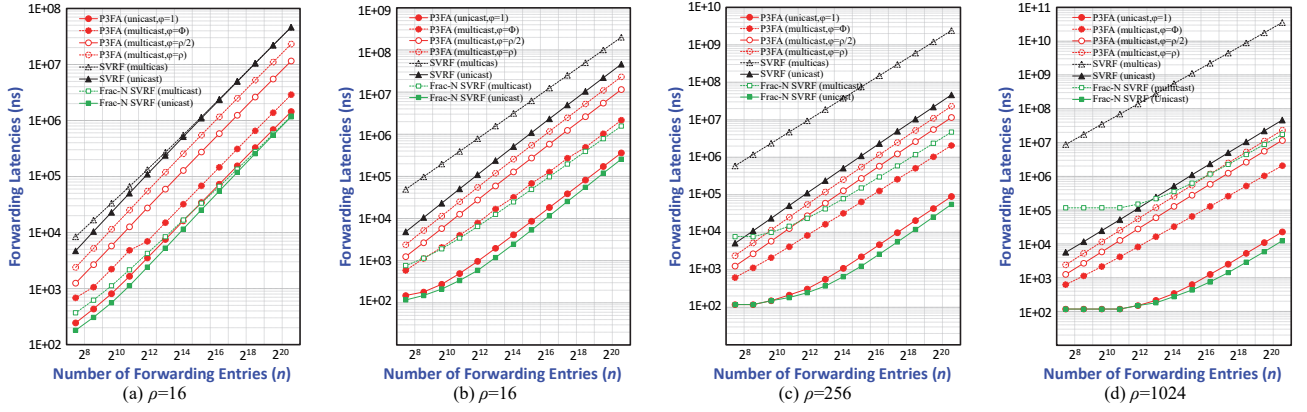

Fig. 3. Forwarding latencies versus number of forwarding entries $n$ in the $\rho$-port PFEs.

### B. Time Efficiency

The most important performance evaluation index of PFE is the forwarding latency, which is mainly affected by the time complexity of the group membership query algorithm. Generally speaking, the forwarding delay will increase as the memory space required by the group membership query algorithm increases, and vice versa. We consider that the hardware accelerators of SVRF/Fractional-$N$ SVRF and P³FA are all constructed by the 2GHz $q$-bit dividers, and each parser/DEMUX/comparer requires 1 cycle. We assume that each memory access requires 10ns through multiple 32-bit data paths, which interconnect the processing and memory units.

The time complexity T of a $q$-bit divider (with 2$q$-bit dividend, and $q$-bit divisors & remainders) is defined by [13], we have

$$T = \left(\left\lceil \frac{\|M_{CP(x)}\|}{q}\right\rceil + 1 + T_O\right)\left(q + 1 + T_O\right)\left(\frac{q}{w}\right), \quad (5)$$

where $\|M_{CP(x)}\|$ denotes the bit-length of a single sub-scalar $M_{CP(x)}$, $T_O$ denotes the overheads between multiple divisions, $w$ denotes the data width of phase shifter built-in the divider. For the SVRF and Fractional-$N$ SVRF schemes, we select $q \geq \|\max(k)\|$ and $\|\min(k)\| \geq \rho$; For the P³FA scheme, we select $q \geq \|\max(k)\|$ and $\|\min(k)\| \geq 2$, where $\|\max(k)\| \leq q$ denotes the maximum bit-length of designated keys.

Fig. 3(a)~(d) show the relationship between the average packet forwarding latency and the number of forwarding entries $n$ at steady state in various $\rho$-port PFEs. In order to compare the time benefits of P³FA under different egress-diversity, we set up 4 different egress-diversity (typically is 1, $\Phi$, $\rho/2$, $\rho$) for control experiments. From the figures, we can first draw one of the most important conclusions, that is, the forwarding latencies of P³FA-based multicast and unicast is much lower than the SVRF whether the port-density is. In addition, with the increase of port-density, the time efficiency of P³FA is more significant. The figures show that even if broadcasting is carried out under the P³FA scheme, its forwarding latency is lower than that of unicast forwarding based on the SVRF. It should be noted that when the port-density is high enough and the number of forwarding entries is low enough, the forwarding latency of the P³FA-based unicasting will be maintained at a low level because it has been bounded within a single instruction time of the 32-bit division.

Although that the time efficiency of Fractional-$N$ SVRF is better than P³FA to a certain extent in specific conditions such as in unicasting, that is because Fractional-$N$ SVRF has made gains at the expense of H/W complexity. Oppositely, the P³FA is more compact than SVRF and Fractional-$N$ SVRF that are many times on its H/W costs.

## IV. Conclusions

This paper discussed a novel forwarding algorithm P³FA based on the prime number theorem to solve the traditional BFs and SVRF's low efficiency and non-scalability problems. Through a series of simulation experiments, we can observe that compared with the SVRF, P³FA has a significant improvement in efficiency and scalability. In terms of space efficiency, when the egress-diversity is low, especially when unicast is the majority, the memory consumptions of P³FA are much smaller than that of SVRF and Fractional-$N$ SVRF. Even with relatively high egress-diversity, the memory consumptions of P³FA may exceed SVRF, but under the high port-density of PFE, we still obtained higher parallelism efficiencies, which greatly improves the time efficiency of PFEs. In brief, our proposed P³FA scheme makes more reasonable use of limited memory resources in PFEs, has a certain degree of scalability, and reduces the forwarding latency effectively.